\documentclass{article}
\usepackage{latexsym}
\usepackage{amssymb}
\usepackage[dvips]{graphics}
\begin{document}
\title{Quantum codes on high-genus surfaces}
\author{Eric Dennis}
\maketitle

\begin{abstract}

An economy of scale is found when storing many qubits in one highly
entangled block of a topological quantum code. The code is defined by
construction of a topologically convoluted 2-d surface and does \emph{not}
work by compressing redundancy in the encoded information.

\end{abstract}

The distinguishing property of a quantum computer is its ability to hold
an amount of data that grows exponentially with the size of the computer.
Entanglement is the main tool for generating and manipulating this data.
And quantum error correcting codes are how we make sure all this goes
according to plan, given finite control over physical components of the
computer. An unexpected connection has been made between quantum codes and
the geometry of 2-d surfaces, rooted in a shared notion of locality
\cite{1}. The resulting geometrical intuition here enables definition of a
code storing many qubits in one large entangled state that exhibits a
novel economy of scale. Each encoded qubit becomes more resistant to
errors as we increase the number of encoded qubits at fixed information
rate (number of encoded qubits per physical qubit).

\begin{figure}[!hp]
\centering
\scalebox{.7}{\includegraphics{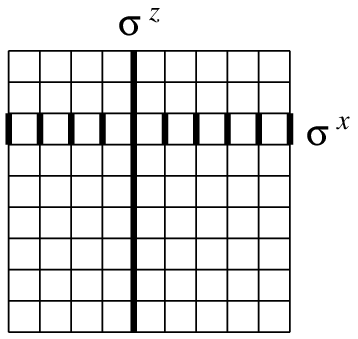}}
\caption{Opposite ends of the lattice are identified to form a torus. The
encoded $Z$ and $X$ operators for qubit one are shown (thick edges).}
\label{torus}
\end{figure}

In one version of the so-called ``topological'' or ``lattice'' quantum
codes, two qubits are encoded in a set of physical qubits identified as
the edges of a $L \times L$ square lattice on the surface of a torus
\cite{1}. The encoded $Z$ operation for qubit number one (two) is given by
any non-contractible loop of $\sigma^z$ operators winding around the torus
in the ``vertical'' (``horizontal'') direction. The encoded $X$ for qubit
one (two) is given by a co-loop (``train-track'') of $\sigma^x$ operators
in the horizontal (vertical) direction; see Fig.\ \ref{torus}. It is known
how to perform an encoded Hadamard transformation and $\pi/2$ phase gate
with these codes. Also, a method exists for performing a Toffoli (C-C-NOT)
gate although with certain complications if very large blocks $\gtrsim
10^9$ are involved \cite{2}.

Errors to physical qubits in a toric code are periodically detected by
``syndrome'' measurements that happen to involve operations which are all
local on the lattice. One performs error correction based on these
measurements, aiming to prevent errors from joining together into long
error chains that might encircle the torus and result in an encoded $X$ or
$Z$ operation being applied unbeknownst to us.

What determines the code's fidelity $1-\epsilon$ in maintaining encoded
information is the length of (number of edges contained in) the shortest
possible non-contractible loops on the lattice, in this case $L$ \cite{3}.
In particular
\begin{equation} \label{epsilon}
\epsilon \sim (p/p_c)^{KL^\beta}
\end{equation}
where $p$ is an error probability for physical qubits, and $p_c$, $\beta$,
and $K$ are parameters depending on the particular error correction 
algorithm used.

$2N$ qubits may be encoded in $N$ separate lattices, each with fidelity
given above. The main result of this paper is that, if instead of $N$
separate lattices we combine them into one large lattice on a high
genus surface constructed by a certain method, the information rate and/or
fidelity may be improved as the number of encoded qubits increases.

As a motivating example, consider joining two $L^\prime \times L^\prime$
toric lattices by removing a $L^\prime/2 \times L^\prime/2$ square from
each and sewing together the perimeters of the resulting square holes; it
is straightforward to define a code on this new lattice preserving the
total number, four, of encoded qubits. Taking $L^{\prime 2} \approx
4L^2/3$, the number of physical qubits is nearly the same as for two
separate $L \times L$ lattices, but the minimum length of non-contractible
loops is now $\approx 2L/\sqrt{3}$, improving the code's fidelity.

\begin{figure}[!hp]
\centering
\scalebox{.7}{\includegraphics{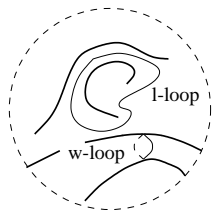}}
\caption{A simple w-loop and simple l-loop are shown on two different
handles.}
\label{loops}
\end{figure}

This suggests, given $N$ separate $L \times L$ toric lattices, we combine
the $2 L^2 N$ physical qubits into one large high-genus surface on which
each torus becomes one handle.  We can increase the code's fidelity by the
following construction.  Cut each handle through its width, along a
``w-loop'' (see Fig.\ \ref{loops}), giving two loose ends per handle. Then
randomly re-pair the set of $2N$ loose ends and rejoin each pair.

Before cutting and rejoining w-loops, the lengths of length-wise
``l-loops'' had been as small as $L$; now the shortest simple l-loops for
a typical handle will have length $\sim LN$. If each handle is made to
encode a qubit with $Z$ operator given by an l-loop, it appears the
chances of a $Z$ error to these encoded qubits has been markedly
diminished. One might like to say this error probability is now
exponentially small in $(LN)^\beta$. But there are more complicated l-type
loops with lengths much smaller than $LN$; each such loop involves many
handles. The minimal l-loops of this kind determine the encoded $Z$ error
probability of a lattice code based on this large surface. Let us first
determine the lengths of minimal l-loops, then symmetrize the construction
of our surface to handle encoded $X$ errors as well.

\begin{figure}[!hp]
\centering
\scalebox{.7}{\includegraphics{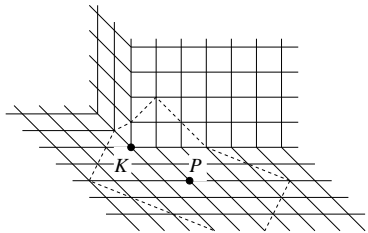}}
\caption{The kink $K$ appears at the base of a handle. A ``circle''
(dashed line) centered at $P$ and containing $K$ is shown.}
\label{kink}
\end{figure}

To characterize the minimal l-loops we will calculate some geometrical
properties of our high-genus surface, on which it will be convenient to
recall the square lattice of qubits. If each handle connects to the
surface through a square patch, the lattice will be locally identical to a
simple square lattice except at the corners of a square patch. Each corner
vertex has valence (number of edges containing it) equal to five not four,
giving five quadrants of 2-d space (see Fig.\ \ref{kink}). The corner
constitutes a kink of negative curvature on the otherwise flat lattice.

Around some vertex $P$ draw a small ``circle,'' \emph{i.e.} a
diamond-shaped locus of vertices equidistant from $P$. As its radius $r$
increases, the circle will encounter handles over which it must climb,
extending out to new places on the surface. From an intrinsic
perspective, the circle merely sees an occasional kink, from which
emanates an extra quadrant of space. Having passed over a kink and
encroached into its extra quadrant, the circle's perimeter will become
larger than it would be apart from the kink. It is not hard to see the
perimeter will contain an additional $r-r_\mathrm{k}$ vertices, where
$r_\mathrm{k}$ is the distance from $P$ to the kink.

As the circle expands, it encounters more kinks and its perimeter grows
even faster. Moreover, all kinks contribute independently to the
perimeter. Assuming a constant density, $8/L^2$ per vertex, of kinks on
the lattice, the perimeter $c(r)$ approximately satisfies the recursion
relation
\begin{equation} \label{rec}
c(r) = 4r + \frac{8}{L^2} \sum_{r_\mathrm{k}=0}^r
c(r_\mathrm{k}) \cdot (r-r_\mathrm{k}), 
\end{equation}
obtained by adding independent contributions from all kinks within the
circle; $c(r)=4r$ would be the result in flat space. The above relation
can be cast as a second order finite difference equation, with initial
conditions, whose solution is approximately
\begin{equation} \label{c(r)}
c(r) = L \sqrt{2} \sinh \left( \sqrt{8} \, r/L \right).
\end{equation}

To obtain the minimal l-loop length for a given handle $H$, consider the
set of open paths of length $r$ and starting at some vertex $P$ around the
base of $H$. As $r$ increases, the circle forming the outer boundary of
this set will be pushed across various handles to random new places on the
surface, gradually filling it up. Once a path encounters the other end of
$H$, it can be closed across $H$ to form an l-loop. The chances there will
be such a path become significant only when the area of our $r$-circle
approaches a significant fraction of the total surface area, $L^2 N$.
Obtaining the circle's area by summing (\ref{c(r)}), this condition is
found to be $r \sim L \log N/\sqrt{8}$, which thus gives the minimal
l-loop length for almost all handles on the surface. As for the other
handles, one can either attempt to re-pair them or simply discard them.

The probability of encoded $Z$ errors associated with l-loops is thus
exponentially small in $(L \log N)^\beta$; however no improvement has been
achieved for $X$ error correction. To symmetrize the above construction,
we simply add an additional cut-and-pair step. Previously we had cut along
a w-loop on each handle and then re-paired all the cuts. Now we cut along
an l-loop, which may be as short as $\sim L \log N$, and randomly re-pair
as before. There is no longer a simple w-loop that can be drawn encircling
a given handle. A w-loop must proceed across many of these cuts before it
can close on itself non-contractibly. Topologically, this new step is
identical to the previous one but with the surface turned inside-out. The
effect of these new cuts and joins is just a doubling of the kink density
to $16/L^2$ in (\ref{rec}), giving the minimal loop length---now for
l-loops and w-loops---as $\sim L \log N / 4$.

This result can be applied either as an improvement to the
$L,N$-dependence of the lattice code's fidelity at fixed information rate,
an improvement to the information rate at fixed fidelity, or as a
simultaneous improvement to both. But there is another parameter to be
considered: the accuracy threshold $p_c$. Indeed, this convoluted surface
topology suggests a greater variety of possible catastrophic error
processes (long error chains) which would tend to worsen the threshold.

Nevertheless, as $L$ is increased at fixed $N$ the error processes
affected by the convoluted topology will only be those involving longer
and longer, hence less and less probable, error chains. For large $L$ the
threshold will then tend back to its original value. Based on analysis
of one quasi-local error correction algorithm \cite{3} with $\beta =
\log_3 2$, it is found that the effect of convoluted topology is to
multiply a bound on the threshold by
\[ 
\sim \prod_{k=1}^{\log_3(L \log N)} 
\left[ \frac{2 (3^k)^2}{a(3^k)} \right]^{(1/3^k)^\beta}
\approx 8 e^{-12(\log N)^{1-\beta}/L^\beta}
\]
where $a(r)$ is the circular area obtained by summing (\ref{c(r)}). This
means that $N$ should not be increased faster than $\log N \sim
L^{\beta/(1-\beta)}$ or else the code's accuracy threshold may be greatly
diminished. Thus the fidelity $1-\epsilon$ from (\ref{epsilon}) will scale
as
\[
-\log\epsilon \sim (L \log N)^\beta \sim L^{\beta/(1-\beta)}
\]

This calculation applies only to the one particular error correction
algorithm for which $\beta$ is fixed as $\log_3 2$; however if we naively
use $\beta = 1$ above, we find there is no restriction on how $N$ scales
with $L$---unbounded gains can be had by increasing $N$ at fixed $L$
without serious damage to the accuracy threshold. In fact one error
correction algorithm has $\beta = 1$, and
explicit consideration of its performance under convoluted topology
corroborates this result. Bounds on the threshold are obtained here by
counting certain classes of paths on the lattice \cite{4}. For instance,
the number of length $r$ paths with given starting point on a flat square
lattice is $4^r$. On our curved lattice, some vertices (the kinks) have
valence 5, so the number of walks is bounded by $v^r$ with $4<v<5$. The
effect of convoluted topology on a threshold bound based on this counting
would be to multiply it by the near-unity factor $4/v$.

It has been shown how to achieve gains in the fidelity and/or efficiency
of storing quantum information by encoding many qubits in one block of a
topological quantum code. The code involves a lattice of qubits on a 2-d
surface of highly convoluted topology. As more encoded qubits are added,
keeping fixed the number of physical qubits per encoded qubit,
asymptotically significant gains are obtained in the code's fidelity. This
is an economy of scale in the error correction hardware independent of any
software gains achieved by compressing redundancy within the encoded
information itself, as in Shannon's coding theorems and their quantum
equivalents \cite{5}, which rely on the encoded qubits' occupation of
``typical'' subspaces in the many-qubit Hilbert space.

One nice feature of the original topological codes is that error
correction operations are local on the lattice; however, this is also a
limitation. The convoluted topology of the above construction, which
effectively destroys the codes' locality, is a way of overcoming this
limitation.

Thanks to U. Madhow and R. Q. Epstein.

\newpage

\end{document}